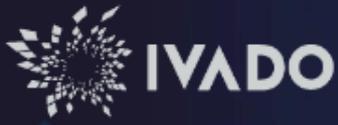 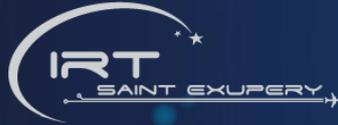 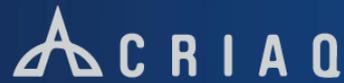 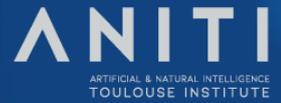

# DEEL
## DEpendable & Explainable Learning

# DATASET DEFINITION STANDARD (DDS)

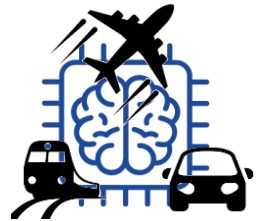

DEEL Certification Workgroup

IRT Saint Exupéry

December 2020

Ref - S079L03T00-018.

**Contributors and affiliations**

- *Cyril Cappi (SNCF / IRT Saint-Exupéry).*
- *Camille Chapdelaine (Safran / IRT Saint-Exupéry),*
- *Laurent Gardes (SNCF / IRT Saint-Exupéry)*
- *Eric Jenn (Thales AVS / IRT Saint-Exupéry),*
- *Baptiste Lefevre (Thales AVS / IRT Saint-Exupéry),*
- *Sylvaine Picard (Safran / IRT Saint-Exupéry),*
- *Thomas Soumarmon (Continental / IRT Saint-Exupéry),*

We would also like to thank the Commissariat Général aux Investissements and the Agence Nationale de la Recherche for their financial support within the framework of the Programme d'Investissement d'Avenir (PIA).

This project also received funding from the Réseau thématique de Recherche Avancée Sciences et Technologies pour l'Aéronautique et l'Espace (RTRA STAe).

**Document history**

| Version | Modifications | Authors | Date |
|---|---|---|---|
| 1.0 | All | See list of contributors. | 2020/01/06 |

**Abstract**

*This document gives a set of guidelines for dataset collection and manipulation in the context of machine learning. It is applicable when the dataset is used to train, validate, or test a ML model.*

*This is a work in progress.*

# Table des matières

# 1  Introduction

This document gives a set of recommendations to build and manipulate the datasets used to develop and/or validate machine learning models such as deep neural networks. This document is one of the 3 documents defined in [1] to ensure the quality of datasets.

This is a work in progress as good practices evolve along with our understanding of machine learning.

The document is divided into three main parts. Section 2 addresses the data collection activity. Section 3 gives recommendations about the annotation process. Finally, Section 4 gives recommendations concerning the breakdown between train, validation, and test datasets.

In each part, we first *define* the desired properties at stake, then we explain the *objectives* targeted to meet the properties, finally we state the *recommendations* to reach these objectives.

# 2  Collecting the data

## 2.1  Data representativeness

### 2.1.1  Definition

In statistics, *a* series of realization of a random variable, or *sample,* is said to be representative of some *population* if it contains key characteristics in proportions similar to the ones of the population [2] [3].

In the machine learning domain, samples are gathered in datasets available for the development of the model, and the *population* corresponds to all the possible observations which can be done in the operational domain.

In the following, we will call data or samples the elements of a dataset that are the realizations of the random variables of interest. Data can be, for instance, pixels, images, or time series…

### 2.1.2  Objectives

During the test phase, the representativeness of the dataset is required to provide a correct evaluation of the actual operational performance of the system. However, during the other development phases (e.g., training), strict representativeness is not required. For instance, some situations may be represented more often than in operation because they are considered to carry more useful information on the behavior to be learnt, or be "more difficult", more critical (e.g. landing and take-off for an aircraft), etc. For this reason, they may deserve to be substantially over-represented in the training phase.

In any case, the dataset used for the development of the model must cover as much as possible the domain of situations that will be encountered by the system in operation.

In the case of image processing, representativeness involve considering the various physical and environmental conditions that may have an impact on the input of the system, including, for instance:

- the objects of interest, including shape, size, color, etc.
- the exposure conditions, including  brightness
- the environmental conditions, including weather, sunlight, temperature, and humidity conditions. A partial list of environmental conditions to check is provided as an example in Appendix.
- the viewing angles
- the distances with respect to the objects of interest

- the presence of occlusions (partial or full)
- the quantity of information about the object of interest (e.g., the number of pixels of the image of the object captured by the sensor).
- The settings of sensors,
- Etc.

A particular aspect of data representativeness is *data currency*, i.e., the *preservation* of representativeness over time. As such, it is not really a property of the dataset building process, but more a property to be ensured by the user of the ML system. For instance, the user of a ML system shall always ensure that the sensors used in the system are actually identical to those used for data collection.

### 2.1.3 Recommendations

**REC 1. The data acquisition chain must be as close as possible to the one that will be used in operation.** ■

In order to make sure that the acquisition process is as close as possible to the one used in the operational conditions of the system, one needs to check that the quality of the data, the acquisition conditions, and the hardware characteristics are similar to the ones in operational conditions. REC 10 covers the case when differences between the data acquisition chain and the operational chain are identified.

**REC 2. System validation[1] should be done on data acquired using the actual acquisition chain.** ■

**REC 3. The Operational Design Domain (ODD) must be defined. The variables characterizing the operational domain (e.g., time of day, brightness, weather conditions, etc.) must be clearly identified.** ■

**REC 4. The datasets used for training and test must be traceable to this ODD.**

**REC 5. The different operational situations must be sufficiently represented. Situations are defined by the type of object of interest (e.g., a specific road sign), and the variation domain of variables characterizing the operational domain (see REC 3).** ■

This recommendation introduces a constraint on the minimum quantity of data that must be collected in order to satisfy the representativity condition. Nevertheless, it is difficult to exactly determine how many data are necessary to ensure a certain precision of the model.

**REC 6. To evaluate the operational performance of the system, the different operating conditions (e.g., brightness, temperature, vibration, etc.) must be represented in the proportions expected during operation.** ■

**REC 7. The necessary and sufficient data shall be maintained to assess that the operational conditions of the ML algorithm comply with the ODD.** ■

---

[1] Validation: confirmation, through the provision of objective evidence, that the requirements for a specific intended use or application have been fulfilled [ISO 9000:2005].

## 2.2 Data traceability

### 2.2.1 Definition

Traceability is a way to maintain a connection between the different development artifacts, including requirement specifications, dataset, models, etc.

### 2.2.2 Objectives

Traceability is a necessary condition to ensure other expected properties such as *reproducibility* (ability to reproduce some result), *reliability* (in particular, ability to know the origin of some data), *confidentiality* (ability to know how private information propagate in the data flow), impact analysis (ability to analyze the impact of some modification), etc.

More generally, traceability is also a means to maintain the history of data [4]. For machine learning, ensuring traceability requires that all the steps for acquiring and building the dataset are perfectly known and verifiable [5].

After their acquisition, data can be subject to several transformations before being used for training, validation, or test. For example, if high-dimensional images are acquired, a common approach to reduce the computational burden of the training of big neural networks is to build the training, validation and test databases with patches taken from these images.

Other transformations, such as unit conversion, compression, or completion [6] can be applied to the data. In some cases, the dataset can also be cleaned, for instance by removing outliers due to some defective or bad-calibrated sensor [6]. Conversely, the dataset can be augmented with simulated data.

At last, in the case where data come from multiple sources, these sources do not necessarily have the same procedures for storing, naming and compressing the data [6]. In this case, when the data are gathered, it is important to make sure that one is always able to identify the source of each data (for instance, by storing the data in different subfolders depending on the source). When they are gathered, if the data needs to be subject to some standardizing (in terms of name, unit or compression format…), then it is important to keep the original data of each source in separate locations.

### 2.2.3 Recommendations

**REC 8.** It shall be possible to regenerate or restore any data used to train or test a model. ∎

This can be achieved, either by saving the data, or by providing means to recompute them, or a combination thereof.

This also makes it possible to restore the data in case of problem during transformation (such as corruption of the data).

**REC 9.** In case of iterative process of data acquisition, a configuration management process shall be used. ∎

If data are collected iteratively, different ML models may be produced from these different sets. It is then very important to be able to identify them.

Data must be traceable from their origin (i.e., parent data and applied transformations). One possibility is to store data coming from different sources into different directories.

## 2.3 Data accuracy and precision

### 2.3.1 Definition

In this section, accuracy and precision refers to the relation between the data collected in the dataset and the data that will be acquired by the system in operation.

Data are accurate if they represent with high fidelity those that will be acquired in operation. Precision refers to the closeness of data related to the same operation situation.

### 2.3.2 Objectives

For image acquisition, accuracy and precision are affected by the fidelity of the image acquisition system used to collect the data used during training and testing to the actual acquisition system used in operation.

For example, it is very important to have an acquisition process for the dataset having the same acquisition frequency, the same resolution, etc. than the final system. If not, real data may appear distorted to the algorithm and the system behavior may become incorrect.

In some cases, it is possible to make the dataset precision closer to the one of the operational data, for example by subsampling. To do so, the acquired dataset must be of higher precision than for the operational system. In that case, it is important to keep a perfect traceability of the transformations as explained in Section REC 6.

Data format may have an impact on data precision, in particular if data are compressed. In this case, it is important that data storage modality do not degrade data precision.

### 2.3.3 Recommendations

**REC 10.** **The differences between the acquisition chain used to acquire the dataset and the acquisition chain used in operation shall be identified and their effect on accuracy and precision shall be estimated.** ■

**REC 11.** **The sources of degradation of accuracy or precision during the dataset elaboration process shall be identified and their effects shall be estimated.** ■

Precision for instance, may be degraded due to compression. As far as possible, lossless compression must be used.

## 2.4 Data reliability

### 2.4.1 Definition

Data are reliable if they are considered credible and relevant for the operational domain.

### 2.4.2 Objectives

Data reliability is crucial to ensure confidence in the model testing and performance assessment. These properties may be ensured by the data sources. In particular, the dataset should be cleaned from any outlier, i.e., any data that is not credible or relevant for the operational domain. Data history should also be transparent to allow for traceability.

### 2.4.3 Recommendations

**REC 12.** **Datasets should be cleaned from outliers.** ■

The notion of "outlier" is defined with respect to the operational domain: *outliers* are data irrelevant for an operation.

**REC 13.   The reliability of the data sources shall be assessed.** ■

## 2.5 Data consistency

### 2.5.1 Definition

Data consistency is defined as the absence of discrepancy between data concerning the same object (e.g., two different birthdays for the same person). Inconsistencies concerning the labels are addressed in Section 3.2.

### 2.5.2 Objectives

Learning or validating a model on the basis of inconsistent (hence, incorrect) data lead to an incorrect behavior or incorrect validation results.

An inconsistency is the violation of some constraint on the dataset. A particular case of constraint concerns repeated data. Note that, independently from their consistency, the rationale for such repeated must be checked with respect to the representativity criterion.

When data come from various sources, the risk of inconsistencies increases [8]. In addition, using different data representations[2] in different sources makes it more difficult to compare the data items, identify the repeated data, and detect the inconsistencies. Therefore, a consistent representation of data shall be favored [9] [10].

In some cases, notably for some image datasets, it may be difficult to express consistency properties. In such case, consistency may only concern annotations (see section 3.2.).

### 2.5.3 Recommendations

**REC 14.   Consistency properties (i.e., constraints on the data) must be expressed. Those properties may concern multiple occurrences of the same data in the dataset, or different data.** ■

**REC 15.   Consistency properties must be verified on the dataset.** ■

**REC 16.   Values of attributes concerning the same "object" must be consistent in all the dataset. Consistency rules shall be expressed and checked over the datasets.** ■

**REC 17.   In a dataset, instances of a same type of data shall use a consistent representation.** ■

This allows for the verification and comparison of each data item and checking for inconsistencies. This recommendation is particularly important for data coming from various sources.

## 2.6 Data integrity

### 2.6.1 Definition

Data integrity is defined as the assurance of the accuracy and integrity of data over its entire entire life cycle" [11].

---

[2] We call the "representation" of data the concrete, physical, manifestation of the data (e.g., in a printed form, in a series of bits, etc.) that is used by a human or a machine to interpret the data.

### 2.6.2 Objectives

In order to avoid unwanted modification of the dataset when it is accessed, an access protocol should be defined. Any modification should be traced in accordance with recommendation given in Section REC 6, and notified to the users.

### 2.6.3 Recommendations

**REC 18.** **Access to the dataset (read, modify) must be granted on a discretionary basis according to the role of each user of the data and its need to carry out his or her activity.** ■

For instance, the users in charge of designing the learning algorithms have access to the final version of the dataset only. Their access rights should be limited to "read access" in order to protect the integrity of the dataset.

**REC 19.** **Any modification of the dataset must be justified, logged, and traced to a user.** ■

**REC 20.** **An access protocol should be defined for users that have a "write" access to the dataset.** ■

**REC 21.** **Users with a "write" access to the dataset should notify modifications of the dataset to other users that may be impacted, and comply with traceability recommendations given in section REC 6.** ■

**REC 22.** **Appropriate measures shall be taken to guarantee the integrity of the datasets (e.g., using some cryptographic hash, etc.).** ■

**REC 23.** **Data integrity must be checked after transmission to a third party.** ■

## 2.7 Unintended bias in data

### 2.7.1 Definition

We will call bias a situation where some features are over- or under- represented in a dataset with respect to the task to be achieved.

In the following example, a "cats and dogs" dataset presents several biases: all cats are represented with a food bowl, all of them are idle and inside a house, and they are essentially facing the camera (two eyes are visible); all dogs are pictured outside over a green background.

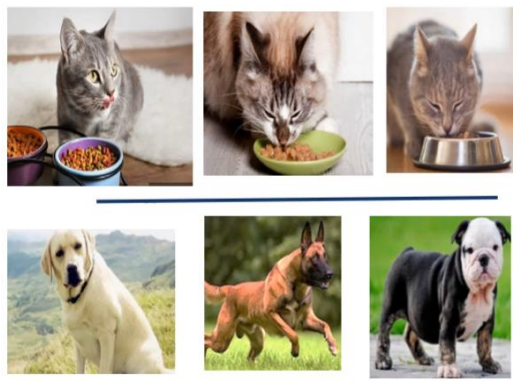

Fig. 2 Example of bias.

Bias in data can present different forms and origins and in general they can be very difficult to detect simply by observing the data. Moreover, bias depends on the task to be performed. For instance, on Figure 2, if the task were to classify cas vs dogs the cats images would be considered as presenting an important bias. However, if it were to detect animals eating, the presence of food bowls would not be considered as a bias (and the dogs images would be considered as outliers).

### 2.7.2 Objectives

As we have seen it, unintended bias minimization is a very complex task. Then to reduce biais occurrence risk two things must be done. First, to specify data collection in a rigourous way taking into account the task to perform and the targeted operational domain. Second, by performing manual and/or automatic bias detection.

### 2.7.3 Recommendations

**REC 24.** **Dataset specifications must be done rigorously before data acquisitions to prevent bias with respect to the intended task and operational context.**

**REC 25.** **Compliance checking must be performed after dataset acquisition.**

## 3 Labelling

### 3.1 Labels accuracy

#### 3.1.1 Definition

Accuracy of labelling is defined as the correctness of the labelling with respect to the true value (or "ground truth') [11], [12].

#### 3.1.2 Objectives

An accurate labelling is necessary to ensure the correctness of the behavior of the system. It is also required to ensure that the performances of the model are correctly evaluated.

For large datasets, an exhaustive verification of labels is often tedious and costly. Therefore, a common usage is to have an *a posteriori* checking of the labels of data for which the worst performances of the model have been observed, *assuming that the labels are mostly correct*.

To evaluate the correctness of a label, the correct label must be known. This is not always an easy task, for example in the case of instance segmentation, where each pixel in an image should be assigned to an object with a class, the contours of objects separating from one object to another can be ambiguous (see Section 3.2). Anyway, it is recommended to work with experts to check the quality of labelling, either before or after having started the work on model development.

#### 3.1.3 Recommendations

**REC 26.** **Experts in the field shall shall check the accuracy of labels on a representative subset.** ∎

### 3.2 Labels consistency

#### 3.2.1 Definition

Consistency refers to the fact that the same label is assigned to the same object of interest when it is represented several times in the data set. In the case where different operators are annotating the same data (multi-labelling), consistency can also be defined by the similarity between the labels provided by the different operators on the same data [13] [14].

### 3.2.2 Objectives

Consistency of labels is needed for convergence in the training of the model.

As explained in Section 3.1, in case of ambiguity, labels for the same object of interest may not match exactly. In that case, labels consistency must be evaluated considering the intended task as for checking labels accuracy.

Here again, experts in the field are recommended to check the consistency in the labels. In the case of multi-labelling, the experts should ensure that the operators assign similar labels for the same data [13].

Ambiguous data are data difficult to label accurately even for an expert. As mentioned in section 3.1, pixel-wise image labeling can be performed on ambiguous data. For instance, it is usually hard to assign a label to the borderline splitting two regions of an image. In that case, labeling is highly uncertain. Ambiguities may also appear when degradations, like glare or occlusions, prevent the good sight of the object of interest. In that case, measures of entropy are not relevant to characterize ambiguity.

### 3.2.3 Recommendations

> **REC 27.** **The same object of interest shall be labelled identically.** ∎

For instance, for the autonomous vehicles' development, the "traffic light" object of interest may occur several times in the data set with various representations. Those representations of the object of interest should be labeled in a consistent manner.

> **REC 28.** **Dataset shall be checked for ambiguities by experts. The presence of ambiguities and the way to address them shall be reported in the annotation procedure (see REC 34).** ∎

The presence of ambiguities shall be captured in the annotation procedure in order to increase the awareness of people in charge of the labelling process.

> **REC 29.** **Consistency and accuracy of labels shall be checked with the same precision.** ∎

> **REC 30.** **Consistency checking and label correction shall be done by experts in the field.** ∎

> **REC 31.** **Ambiguous data should preferably be labelled manually, not automatically.** ∎

This makes it easier to discuss with the experts on the correct label that should be used.

> **REC 32.** **If the labelling process is done automatically, the whole automatic workflow shall be checked to see how ambiguous data are processed and the quality of the workflow must be assessed.** ∎

## 3.3 Absence of Annotation bias

Data bias can be present in the data collection or in its annotations. We focus in this paragraph on annotation bias [14].

### 3.3.1 Definition

Annotation bias is defined as a distortion of the annotation process.

Annotation biases appear when the annotation is not only dependent on the information present in the data. That means it takes into account other elements internal to the annotation process. These elements can lead to annotation errors. Annotation bias may correspond to cognitive bias due to repetitive processing of strongly similar data. Another annotation bias can come from a lack of annotation guidelines.

Annotation biases are subsequent to annotation. All internal elements of the annotation process should be taken into account in order to prevent this [15] [16].

### 3.3.2 Objectives

Recommendations aim at reducing annotation bias.

### 3.3.3 Recommendations

**REC 33.** **The ability of the people in charge of the annotation process must be assessed.** ■

**REC 34.** **Annotation instructions must be defined and validated by experts in the field.** ■

**REC 35.** **The quality of the annotation process shall be assessed by verifiying a statistically significant sample of the annotated data.** ■

For instance, one must be vigilant about cognitive biases that can lead to annotation biases. ■

This recommendation is not really easy to control, since much of its compliance depends on the annotator himself. To help the annotator, and thus better insure bias avoidance, we can apply the recommendation REC 36.

**REC 36.** **A specific checking shall be done when a series of data have the same label.** ■

This recommendation aims at detecting a possible cognitive bias of the annotator.

**REC 37.** **Data shall be assigned to annotators in a random way.** ■

This recommendation is a good practice because almost similar data are often stored nearby in the database. One possibility to avoid cognitive bias is to distribute the data to the annotators in a disordered manner with respect to their original places of storage.

**REC 38.** **Annotations shall be traceable to the person that did the annotation.** ■

This can enable, in the learning process, to identify any cognitive biases that some annotators have incidentally introduced. To do this, the table or the program mentioned for recommendation REC 36 can be used.

## 4 Train, Validation, and Test datasets.

### 4.1 Building the datasets

Data must be separated into three distinct datasets, known as training (or learning), validation and testing.

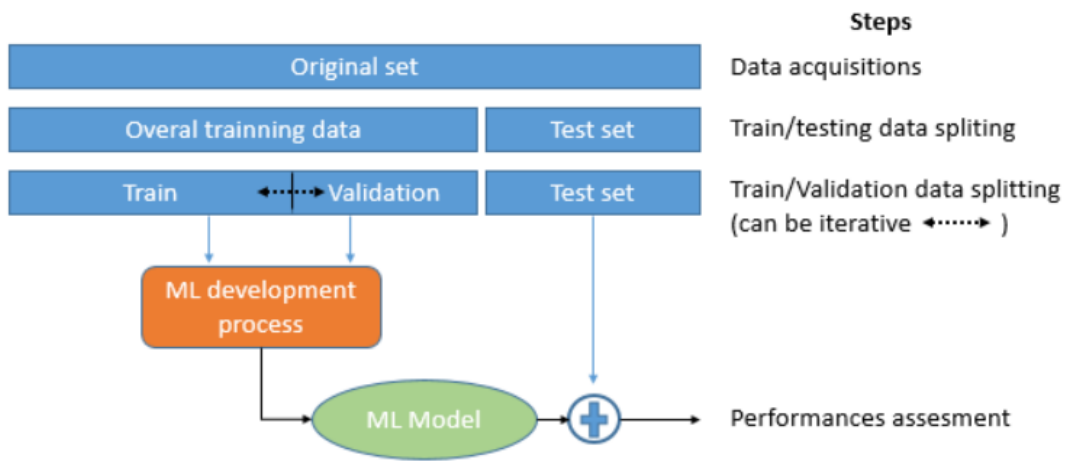

Fig. 2 Data life-circle in ML development process.

### 4.1.1 Definitions

The training dataset is used to set the model parameters. This calculation is performed by training algorithms, several of which are presented in chapter 8 of reference [15].

The validation dataset is used during training to check that the model generalizes to a different dataset than the training set. It is also used to set the hyper-parameters of the learning algorithm, such as the learning rate.

Once the model has been trained and validated, the test dataset is used to evaluate the performances of the model. It is worth noting that these tests can cover operational performances or robustness tests. However, in the following, the purpose of the testing dataset largely focuses on operational performance evaluation.

### 4.1.2 Objectives

The test dataset is of crucial importance for evaluating the performance of the model. The constitution of the test dataset is therefore delicate and must be done in a thoughtful manner.

For the evaluation of operational performances, the test dataset shall present data in the operational data distribution. Additional tests may be added to cover situations deemed more difficult to classify by experts or other particular cases. In this sense, the test makes it possible to verify the robustness of the model, in addition to its overall performance. These test cases may for instance correspond to highly noisy or degraded data (e.g. due to acquisition failure). Difficult cases can also include images with a lack of visibility.

Data can also be considered as difficult depending on the behavior of the model itself (for example, in the case of a rare situation, or a blurred decision boundary).

### 4.1.3 Recommendations

**REC 39.** **The train set and test set must not overlap.** ∎

**REC 40.** **The test dataset must be kept secret until the model has been fully validated.** ∎

This separation is a fundamental condition for correct model performance assessment. Consequently, communication must be marked out between the teams constituting respectively the train and test datasets.

**REC 41.** The test team must define the level of information it can convey to the train team without the risk of introducing bias during the test. ∎

## 4.2 Independence between datasets

### 4.2.1 Definition

In the context of machine learning, independence between datasets means that there is no way to predict the data of a dataset from the observation of another dataset.

### 4.2.2 Objectives

Independence between data is necessary to be able to apply standard machine learning algorithms [17]. If this independence cannot be guaranteed, particular training methods and/or performances metrics must be used [18] [19].

Within the training, validation and test dataset, as well as between these datasets, independence between data must be ensured.

From the test point of view, putting the same image in train and test sets biases the evaluation of the model's performance. This is called "data leakage".

In order to avoid these leaks during data acquisition and manipulation, protocols must pay a particular concern to this question. If data augmentation techniques have been used, the augmented data should be included in the same dataset as their original data.

As an illustration, in the case of video sequences, the correlation between the consecutive frames is high. Then if the sizes of the datasets are limited, this high correlation leads to redundant information that should be avoided. Indeed, this redundancy limits the variety of the observed data and then can have a bad impact on representativeness and operational domain representation. Moreover it can introduce bias.

### 4.2.3 Recommendations

**REC 42.** Automatic detection of bias must be performed on training data. ∎

Bias in the data can be present on purpose (e.g. to perform some robustness assessment). However, unintended bias must be detected in training set.

**REC 43.** Data redundancy must be avoided between the train and test datasets in order not to bias the evaluation of the model's performance. ∎

For images, if the images are taken from videos, care must be taken not to include several consecutive images from the same video. Even if the data are not the same, they are too close to each other to be put in different sets.

**REC 44. REC 41.** The size of the test set shall satisfy the following relation

$$\mathbb{P}\left(p \leq \widehat{p}_n + \sqrt{\frac{2\widehat{p}_n}{n} \ln\left(\frac{1}{\delta}\right)} + \frac{2}{n} \ln\left(\frac{1}{\delta}\right)\right) \geq 1 - \delta$$

> **Where $\delta$ is the desired confidence level, $\widehat{p}_n$ is the rate of incorrect classification observed on the test set, $p$ is the actual unknown probability of incorrect classification, and $n$ is the size of the test set.**
>
> **This relation is applicable under the assumption that the samples are i.i.d and drawn according to the target distribution. The target distribution depends on the evaluation context.** ∎

This formula states the size of the test set ($n$) necessary to bound the actual probability of error with a confidence of 1- $\delta$. The larger $n$ is, the tighter the bound is (see [18] [21] for more details about the derivation of the formula).

## 5 Conclusion

We have presented a set of recommendations aimed at ensuring the quality of the datasets used to build machine-learning models. This set is still partial, and, in particular, the issue of data representativeness would require more investigations. However, we think that confidence will be achieved by combining three approaches contributing to representativeness: ensuring representativity by an appropriate building of the datasets, constraining the operational domain by appropriate operational rules (when possible), and preventing using the models on data that are out of this distribution.

# Appendix
# Environmental conditions

| Domain | Variable / effect | Quantification/Quantization |
|---|---|---|
| **Weather** | Fog | Level in {strong, low, no} <br><br> *Information may be (partially) obtained from meteorological data.* |
| | Heat haze (refraction due to the uneven temperature of air) | Level in {strong, low, no} <br> *Information may be (partially) obtained from meteorological data.* |
| | Rain | Level in {strong, low, no} <br> *Information may be (partially) obtained from meteorological data.* |
| | Snow | Level in {strong, low, no} <br> *Information may be (partially) obtained from meteorological data.* |
| | Sun | Level in { very cloudy, cloudy, clear} <br> *Cloud cover can be measured by the fraction of the sky obscured by clouds.* |
| **Light conditions** | Sun position | Phase in {dawn, zenith, crepuscule} <br> *Sun position could be measured by its elevation computed from the ephemerids.* |
| | Relative position of the sun / train | Relative position in {front, back, side} <br> *Relative position could be expressed by the azimuth of the sun with respect to the camera. This quantity can be computed from the ephemerids.* |
| | Ambient light level | Level in {high, low} <br> *The light level could be measured by a luxmeter.* |
| | Light gradient *Effect on the sensibility due to the pupillary light reflex (and, possibly, saturation of the retina?)* | Gradient in {high, low, no} <br> *The gradient could be measured by a luxmeter (and derivation / time)* |
| | Other light sources | Light sources in { car light, work in progress signaling, laser } |
| **Railway signal** | Graffiti | Level in {yes, no} |

| | Broken signal (e.g., rupture of the white border) | Level in {yes, no} |
|---|---|---|
| | Position | Position in {on gallows, on pole} |
| | Distance to the train | Distance in {close, medium, large} |
| | Type of signal | Signal type in {unique, combined, multiple (for several tracks) |
| **Location** | Location  *Effect on the background of the image. In cities, possibility to have traffic light, urban lighting, etc.* | Location in {station, city, country side, mountainous area, tunnel} |
| | | *This information can be obtained from the location of the train and the cartography.* |
| **Occlusions** | Occluding element | Occluding element in {catenary poles, signaling poles, bridges, vegetation (including leaves, pollens,…), crossing train, birds, dirt (on the windshield), plastic bags, flying tarpaulin,…) |
| | | *This information may be obtained by comparing images in a sequence.* |

*Table 1 : Environmental Conditions.*